\documentstyle[12pt]{article}
\begin{document}
\baselineskip=20 pt
\def\delmunul{\Delta_{\mu\nu}}
\def\bmunul{B_{\mu\nu}}
\def\e{\eta}
\def\k{\kappa}
\def\s{\sigma}
\def\m{\mu}
\def\n{\nu}
\def\r{\rho}
\def\a{\alpha}
\def\b{\beta}
\def\g{\gamma}
\def\d{\delta}
\def\ep{\epsilon}
\def\L{\Lambda}
\def\mphi{m_{\phi}}
\def\vphi{<\phi>}
\def\bea{\begin{eqnarray}}
\def\eea{\end{eqnarray}}
\begin{center}
{\Large{\bf Tadpole diagrams due to KK modes of graviton

 and radion do not contribute to the~ $\rho$ parameter}}
\end{center}

\vskip 10pT
\begin{center}
{\large\sl \bf{Prasanta Das}~\footnote{E-mail: pdas@iitk.ac.in}
}
\vskip  5pT
{\rm
Department of Physics, Indian Institute of Technology, \\
Kanpur 208 016, India.} \\
\end{center}

\begin{center}
{\large\sl  \bf{Uma Mahanta}~\footnote{E-mail:mahanta@mri.ernet.in}
}
\vskip 5pT
{\rm
Mehta Research Institute, \\
Chhatnag Road, Jhusi
Allahabad-211019, India .}\\
\end{center}  

\vspace*{0.02in}

\begin{center}
{\bf Abstract}
\end{center}
In this brief report we show that tadpole diagrams due to KK modes
of the graviton and radion
 do not contribute to the vectorial isospin breaking
$\rho$ parameter. Our result is fairly general and it holds both for
ADD and RS type scenarios. It shows that the difference between the results
of different publications on this subject is not due to some of them
not having considered tadpole diagrams in estimating the contribution
to the $\rho$ parameter.

\newpage
Recently there has been a lot of activity in putting bounds on the
unknown parameters of models of extra dimensions
[\cite{ADD}, \cite{RS}]. The  oblique electrweak
(EW) parameters S, T and U in particular have been used to put bounds
on the unknown parameters [\cite{DR}, \cite{HMZ}, \cite{CGK}] of the
model. However the
analysis of these
papers differ considerably among themselves and so also their result. For
example in Ref. 3 the tadpole diagrams due to Kaluza-Klein (KK) modes of
the  graviton and the radion were considered in estimating  the
contribution of sub-millimeter size extra dimensions to the $\rho$
parameter. However no such tadpole diagrams were considered in Ref. 4 to
estimate the 
$\rho$ parameter in the same model. The controversy between the two
papers is particularly serious because of two reasons: i) firstly tadpole
diagrams are sometimes necessary to rastore gauge invariance of the final
result and ii) secondly
the bounds given in Ref. 3 are much stronger than those of Ref. 4.
Further in a footnote of Ref. 3 it has been mentioned that 
the results of Ref. 4 should not be trusted since 
the authors of that paper do not consider
the contribution of tadpole diagrams to the $\r$ parameter.
It is therefore crucial to find out if the difference between the results
of Refs. 3 and 4 really originates from tadpole diagrams.

In this brief report we shall explicitly show that the tadpole diagrams
due to KK modes of graviton and radion
 do not contribute to the $\rho$ parameter. Our result
implies that tadpole diagrams are $SU(2)_v$ symmetric. It also shows 
that the difference between the results of Ref. 3 and Ref. 4 does not
originate from tadpole diagrams but from somewhere else e.g. in the treatment
of infrared divergences.

Let us first consider the contribution of a KK graviton tadpole to the
$\rho$ parameter. Let $\Delta_{\mu\nu,\a\b}(k=0)$ denote the propagator
for the KK graviton shown in Fig1. We then have  \cite{HLZ}

\bea
i\Delta_{\mu\nu,\a\b} (k=0)=- i{B_{\mu\nu,\a\b} (k=0)\over
m^2_n}
\eea
where
\bea
B_{\mu\nu,\a\b} (k=0)= \e_{\mu\a}\e_{\nu\b}+
\e_{\mu\b}\e_{\nu\a} - {2\over 3}\e_{\mu\nu}\e_{\a\b}
\eea
and $m_n$ is the mass of the $n^{th}$ KK mode of the graviton.

The Feynman rule for the graviton vertex with two gauge bosons
 in unitary gauge ($\xi=\infty$) is given by
\cite{HLZ}
\bea
-i{\k\over 2}[(m^2_v-q^2)(\e_{\mu\rho}\e_{\nu\sigma}+\e_{\mu\sigma}
\e_{\nu\rho}-\e_{\mu\nu}\e_{\rho\sigma})+D_{\mu\nu , \rho\sigma }(q, -q)]
\eea
where
\bea
{D_{\mu\nu ,\rho\sigma }(q, -q) = -\e_{\mu\nu}q_{\sigma}q_{\rho}
+\left[\e_{\mu\sigma}q_{\nu}q_{\rho}+\e_{\mu\rho}q_{\sigma}q_{\nu}
+\e_{\nu\sigma}q_{\mu}q_{\rho}+\e_{\nu\rho}q_{\mu}q_{\sigma}-
2\e_{\rho\sigma}q_{\mu}q_{\nu}\right]}
\eea
and $m_v$ is the mass of the EW vector boson (W or Z) under consideration.
Hence
\bea
i\Pi^{tadpole}_{\rho\sigma}(q) = {\k\over 2}[(m_v^2-q^2)
(\e_{\m\r}\e_{\n\s}+\e_{\m\s}\e_{\n\r}-\e_{\m\n}\e_{\r\s})+
D_{\m\n, \r\s}(q, -q)] \noindent
\nonumber \\
\times {1\over m^2_n}B_{\m\n\a\b}(k=0)\int {d^4l\over (2\pi)^4}{i \k\over
2} V^{\a\b\g\d}(l)i P_{\g\d}(l)
\eea
Here $i {\frac{\k}{2}} V^{\a\b\g\d}(l)$ stands for
the graviton vertex with 
the closed loop and $i P_{\g\d}(l)$ stands for the propagator of the 
particle circulating in the loop shown in Fig.1. The vertex function
$V^{\a\b\g\d}(l)$ is symmetric in $(\a, \b)$ and $(\g, \d)$ separately.
The tensorial structure
assumed for the vertex $V^{\a\b, \g\d}(l)$ and the propagator $P_{\g\d}(l)$
holds if the particle circulating in the loop is any SM particle or a radion.
However if the particle circulating in the loop is a KK graviton then the 
form of the vertex and the propagator must be generalized to 
$V^{\a\b\g\d}(l)$ and $P_{\g\d}(l)$. Our result will however be 
valid irrespective of the kind of particle that circulates in the loop.
Consider the loop integral $\int {d^4 l\over (2\pi )^4}V^{\a\b, \g\d}(l) 
P_{\g\d}(l)$.  Since there is no external momentum flowing into the loop,
the only second rank symmetric tensor that can be formed after the loop
integration has been done must be proportional to $\e^{\a\b}$.
Hence we can write   $\int {d^4 l\over (2\pi )^4}V^{\a\b, \g\d}(l) 
P_{\g\d}(l)=i \e^{\a\b}F(m^2_p, \L^2)$. Here $m_p$ is the mass of 
the particle circulating in the loop and $\L$ is the ultraviolet
momentum cut off.
We then have

\bea
\Pi^{tadpole}_{\r\s}(q)=({\k\over 2})^2 \left[m^2_v(\e_{\m\r}
\e_{\n\s}+\e_{\m\s}\e_{\n\r}-\e_{\m\n}\e_{\r\s})
+ D_{\m\n\r\s}(q, -q)\right]{B_{\m\n\a\b}(k=0)\over m^2_n}\e^{\a\b}
\noindent
\nonumber \\
\times \sum_p F(m^2_p, \L^2)
={\k^2\over 3}{m^2_v\over m^2_n}\e_{\r\s}\sum_p F(m^2_p, \L^2)
\eea
The summation over p takes care of all the particles circulating 
in the closed loop.
The function $F(m^2_p, \L^2)$ occurs both for W and Z boson self 
energies.
It then follows that 
$${\Pi_{ww}(0)^{tadpole}\over m_w^2}-
{\Pi_{zz}(0)^{tadpole}\over m_z^2}=0$$
Hence graviton tadpole diagrams
do not contribute to the $\r$ parameter. The proof that radion tadpole
diagrams also do not contribute to the $\r$ parameter is much easier.
The Feynman rule for the vertex function of a radion with two gauge
bosons is 
$$2 im^2_v \e_{\r\s}{1\over <\phi>}$$ 
where $<\phi>$ is the radion
vev and $m_{\phi}$ its mass. The radion line shown in Fig 2 does not
carry any momentum. The Feynman rule for it is therefore $-{i\over
 m_{\phi}^2}$. Finally let $G (m^2_p, \L^2)$ denote the result of doing 
the loop integration in Fig. 2. For the radion tadpole diagram  the gauge
boson vacuum polarization tensor becomes

$$i\Pi^{tadpole}_{\r\s}(q)=i{m^2_v\over \mphi^2 \vphi^2}\e_{\r\s}\sum_p
G(m^2_p, \L^2)$$
It then follows that  
$${\Large{\Pi_{ww}(0)^{tadpole}\over m_w^2}~-~ {\Pi_{zz}(0)^{tadpole}\over
m_z^2}=0}$$ 
for radion tadpole diagrams also.
Hence radion and graviton tadpole diagrams do not contribute to the
$\r$ prameter both in ADD and RS scenarios. We can therefore conclude
that the difference between the results of Ref. 3 (strong constraints)
and Ref. 4 (weak constraints) does not arise from tadpole diagrams
which were considered in the former but ignored in the later.

\newpage

\begin{figure}[htb]
\begin{center}
\vspace*{5.5in}
      \relax\noindent\hskip -5.4in\relax{\includegraphics{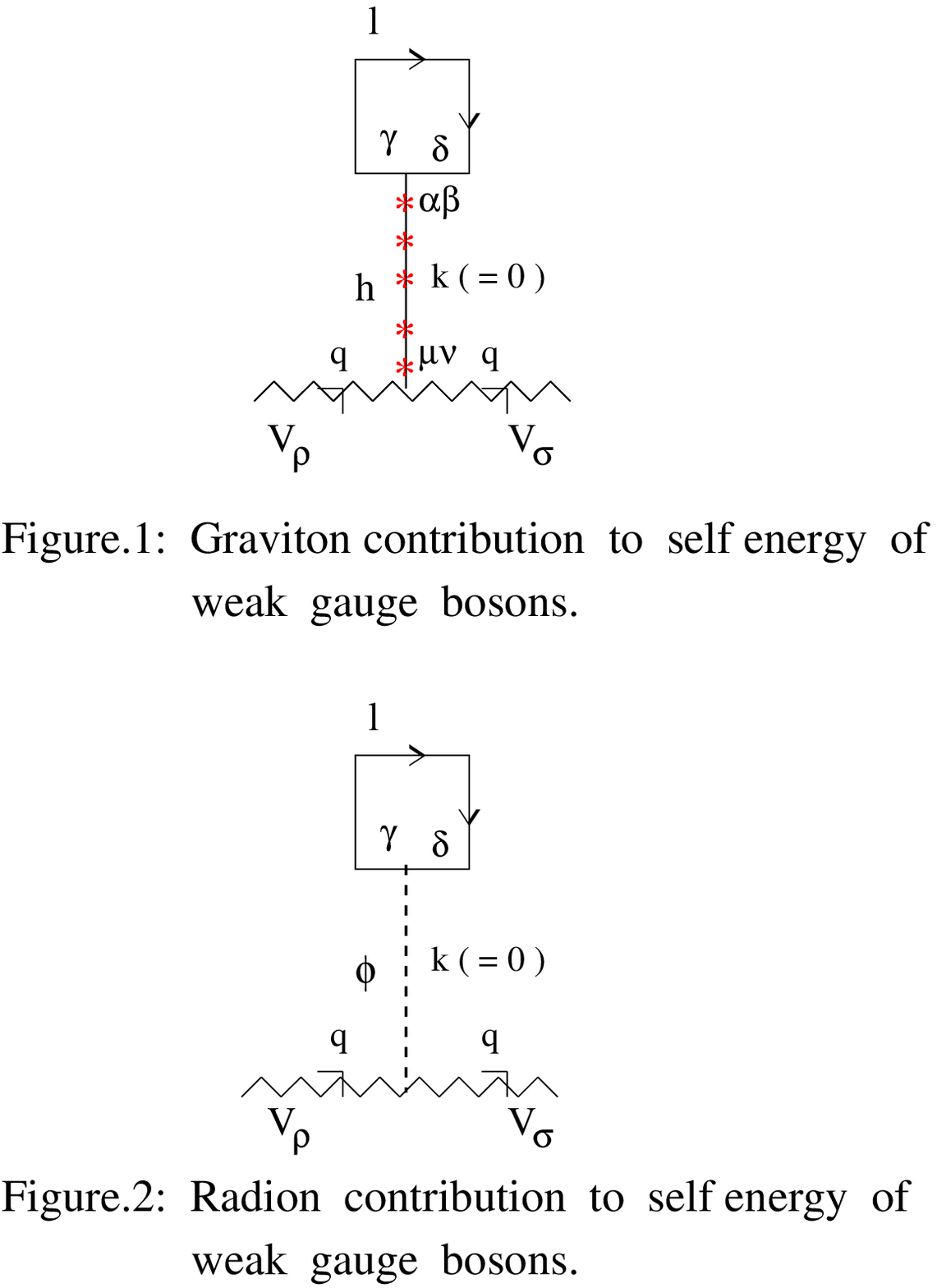}} 
\end{center}
\end{figure}  
\end{document}